\documentclass[twocolumn,showpacs,preprintnumbers,amsmath,amssymb]{revtex4} 

\usepackage{graphicx}
\usepackage{dcolumn}
\usepackage{bm}

\begin{document}

\title{Surface-induced near-field scaling in the Knudsen layer of a rarefied gas}
\author{R. R. Gazizulin, O. Maillet, X. Zhou, A. Maldonado Cid, O. Bourgeois, and E. Collin}

\address{Universit\'e Grenoble Alpes, \\
Institut N\'eel CNRS, \\
25 rue des Martyrs, BP 166, 38042 Grenoble Cedex 9, France \\
}

\date{\today}

\begin{abstract}

We report on experiments performed {\it within} the Knudsen boundary layer of a low-pressure gas. 
The non-invasive probe we use is a suspended nano-electro-mechanical string (NEMS), which interacts with $^4$He gas at cryogenic temperatures. When the pressure $P$ is decreased, a reduction of the damping force below molecular friction $\propto P$ had been first reported in Phys. Rev. Lett. {\bf 113}, 136101 (2014) and never reproduced since.
We demonstrate that this effect is independent of geometry, but dependent on temperature. 
Within the framework of kinetic theory, this reduction is interpreted as a {\it rarefaction phenomenon},  carried through the boundary layer by a deviation from the usual Maxwell-Boltzmann equilibrium distribution induced by surface scattering.
Adsorbed atoms are shown to play a key role in the process, which explains why room temperature data fail to reproduce it. 

\end{abstract}


\maketitle

%
%

A low density gas is statistically described by the well-known Boltzmann kinetic theory \cite{Carlo,Chapman,grad}. The equilibrium state in the bulk, that is the distribution function $f$ characterizing the molecular motion that cancels the collision integral $Q(f,f)$, is simply the well known Maxwell-Boltzmann (MB) distribution $f_0$. 

In any physical situation this equilibrium is imposed by {\it boundary conditions}: the gas is at temperature $T_0$, and pressure $P_0$ enforced by e.g. the walls of a container. The interaction between gas particles and the solid surface is thus crucial, even in such a simple situation; in more complex cases where for instance a gas flow is forced near the wall, the presence of the interface generates unique features like slippage and temperature jumps \cite{Carlo,patterson,Sone,lockerby,microChevrier}. All of this happens in a layer of thickness a few mean-free-paths $\lambda$, the so-called Knudsen boundary layer.

These features are essential in aeronautics and in the expanding field of micro/nano-fluidics \cite{favero,Bocquet,microChevrier}; but their accurate modeling remains a challenge, even using today's numerical computational capabilities \cite{dadzie,sader,saderII,scientreports}. 
Already in the early days of the kinetic theory development, Maxwell had noticed the importance and difficulty represented by the boundary problem \cite{Maxwell}; his discussion of molecular reflections (introducing an accommodation parameter $p$) is still valuable today. 

The problem is indeed nontrivial, since the scattering mechanism on the wall depends intimately on details of complex surface physics phenomena like adsorption and evaporation of molecules. Besides, this introduces a strong asymmetry between incoming particles reaching the wall with the statistical characteristics of the gas, while escaping particles carry properties defined by the solid body \cite{Carlo,patterson,Sone,dadzie,sader}.

In the present Letter we report on measurements performed within the boundary layer by means of a high quality nano-electro-mechanical device (NEMS). We use $^4$He gas at cryogenic temperatures, which is an almost-ideal gas with tabulated properties \cite{NIST}. When the pressure $P$ is sufficiently low, such that the mean-free-path of atoms is sufficiently long, we measure a decrease of the gas damping below the well-known molecular law $\Delta f_{molec.} \propto P$. Comparing different devices and measurements performed at different temperatures, we show that this anomalous decrease is consistent with a {\it reduced gas density} within the boundary layer. We can justify it as a deviation from the standard Maxwell-Boltmzann equilibrium distribution induced by the presence of the wall and its adsored atoms: a near-field effect propagated from the actual boundary scattering mechanisms that decays within the gas over a few mean-free-paths $\lambda$.

\begin{figure}\vspace{-1.5cm}
\includegraphics[height=12 cm]{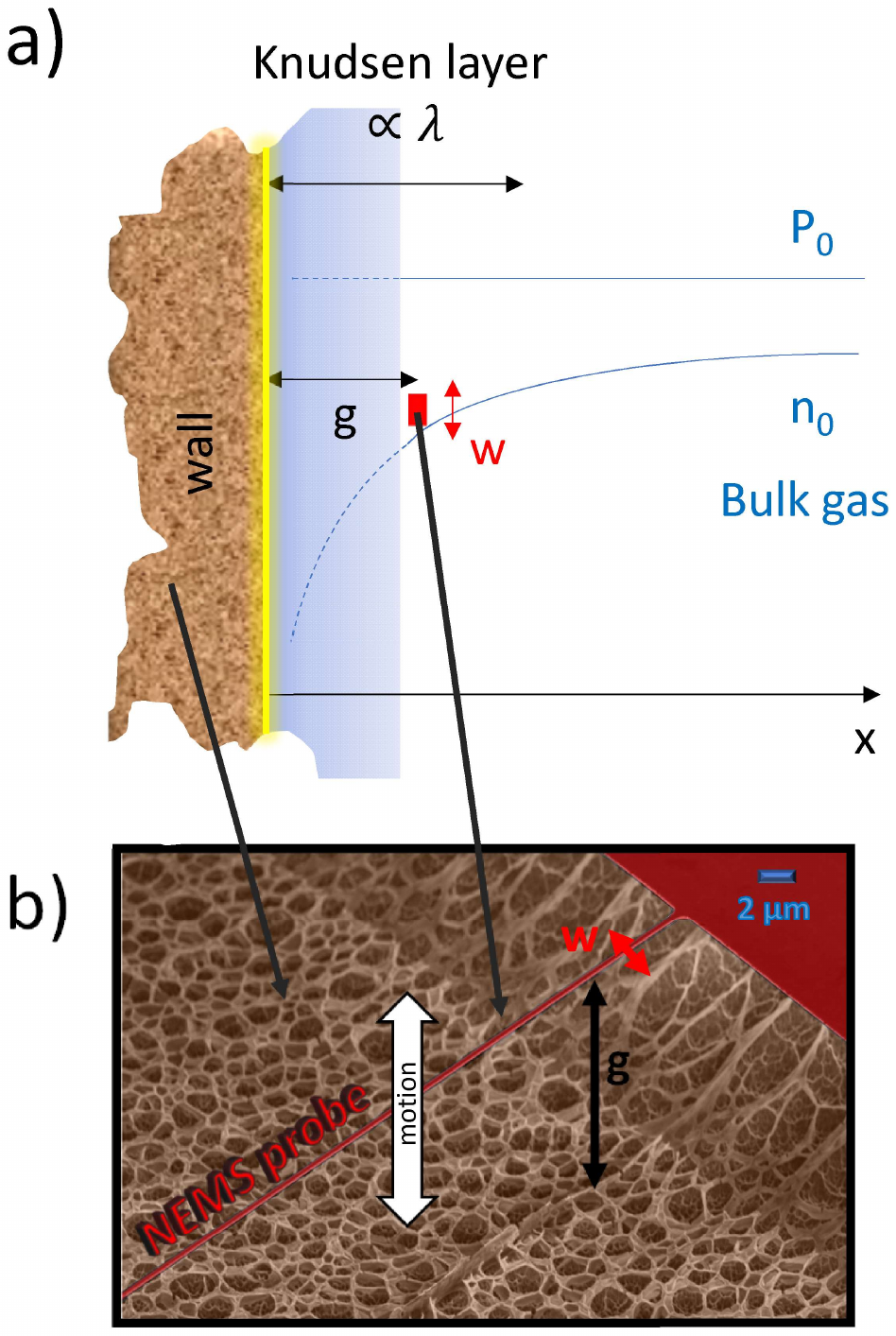}
\vspace{-1.5cm}\caption{\label{figure1} (Color online) {\bf a)}: Schematic of the experimental situation with in brown the container's surface (on left), in light yellow the adsorbed atoms and then the gas (bulk on the far right). The pressure $P_0$ remains constant while the density $n(x)$ drops from $n_0$ as we approach the wall, within a thickness of order a few mean-free-paths $\lambda$ (Knudsen layer). 
Our NEMS local probe of width $w$ lies a distance $g$ from the surface (red rectangle). Close to the wall, macroscopic theories based on simple expansions fail to describe the physics, and delicate microscopic modeling/numerical simulations are required (shaded area). {\bf b)}: Actual device used in this work (SEM false color image), which oscillatory motion is orthogonal to the wall (arrow). Note the spongy nature of the interface.}
\end{figure}


In Fig. \ref{figure1} we show a schematic of the setup and a Scanning Electron Micrograph (SEM) of the device. 
It consists of a high-stress silicon-nitride NEMS beam of $L=150~\mu$m length, width $w=300~$nm and thickness $e=100~$nm. A 30$~$nm thick aluminum layer has been deposited on top for electrical contacts \cite{JLTPKunal}. The moving structure is suspended at a distance $g=20~\mu$m above the bottom of the etched chip. Its motion is actuated and detected through the magnetomotive scheme \cite{SensActCleland,RSIus}:
a current $I_0 \cos(\omega t)$ oscillating at frequency $\omega$ close to the first out-of-plane flexure resonance $\omega_0$ is fed in the metallic layer while the structure resides in an in-plane magnetic field $B_0$ orthogonal to the beam. 
The resulting Lorentz force $F_0$ generates the motion while the voltage induced by the cut magnetic flux $V_0$ is detected by means of a Stanford\copyright $~$SR844 lock-in amplifier. Details on the scheme and calibration can be found in Ref. \cite{RSIus} and Supplementary Material \cite{SM}.

The device is glued on a copper plate, which is mounted in a chamber placed inside a $^4$He cryostat. 
The pressure is measured at room temperature using a Baratron\copyright $~$pressure gauge. $^4$He gas was added by small portions to the cell from the evaporation of a dewar connected through a needle valve. The temperature is lowered from 4.2$~$K down to 1.3$~$K by pumping on the $^4$He bath of the cryostat. It is regulated up to 20$~$K by means of a heater attached to the copper sample holder. The temperature is measured from the other side with a calibrated carbon resistor connected to a bridge. More details can be found in Supplementary Material \cite{SM}.

\begin{figure}[h!]
\includegraphics[height=7.5 cm]{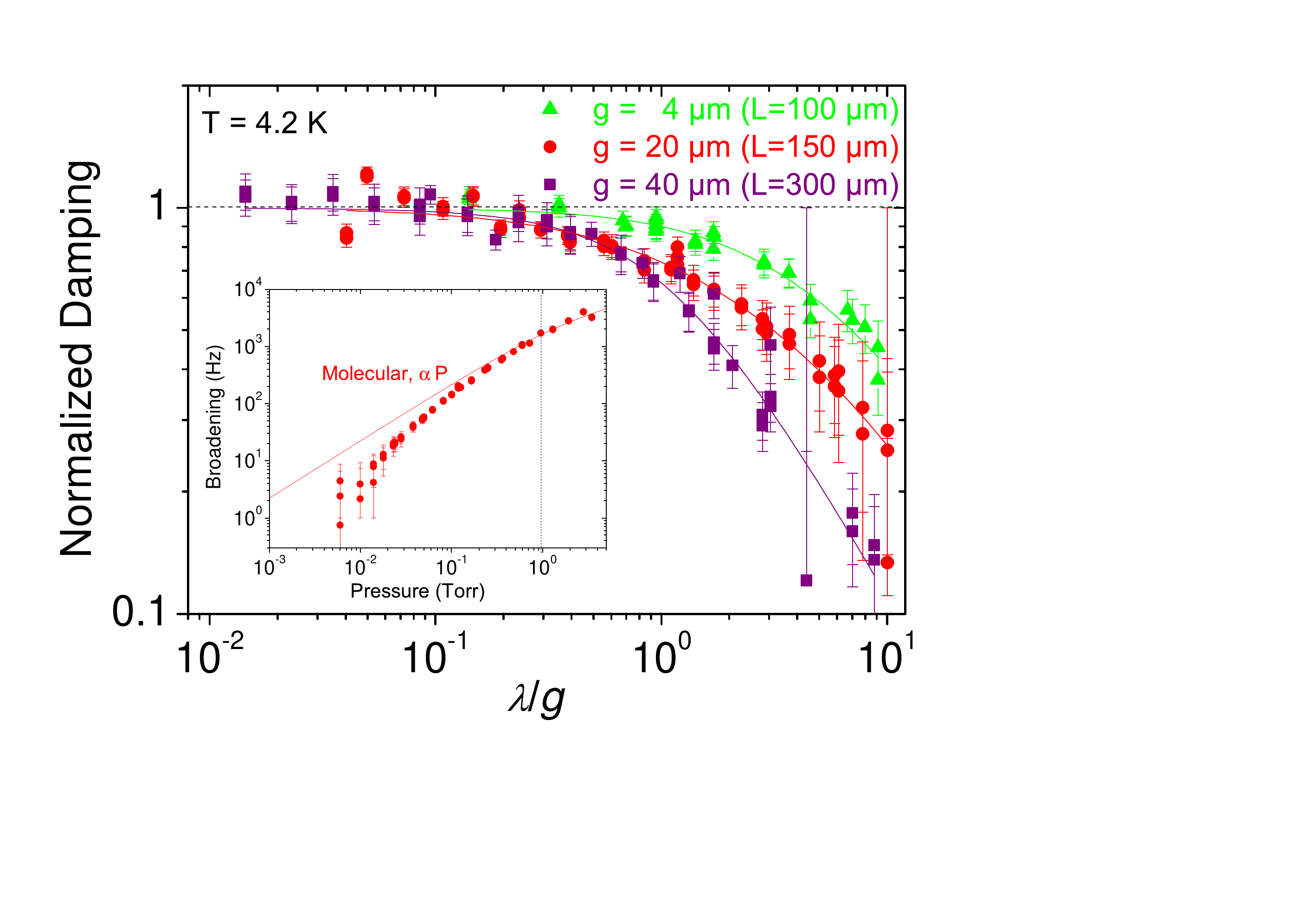}
\vspace{-1.5cm}\caption{\label{figure2} (Color online) Main: Deviation form molecular damping (proportional to $P$) measured at low pressures, for 3 devices at 4.2$~$K as a function of the Knudsen number $\lambda/g$. The lines are fits to Eq. (\ref{fit}), see text. Inset: Raw data for the 150$~\mu$m long device (intrinsic damping subtracted); the full line is the standard molecular damping expression, and the vertical dashed line shows the cross-over to Navier-Stokes damping (see text).}
\end{figure}

We first perform measurements at 4.2$~$K as a function of pressure. Subtracting the intrinsic damping of the mode (about 100$~$Hz), we plot in Fig. \ref{figure2} inset the broadening $\Delta f$ of the resonance as a function of $P$. 
The drive has been kept low enough to be in the linear regime, and the resonance is a Lorentzian peaked around 1.65$~$MHz. No particular nonlinear damping has been noticed in the measurements. 
When the pressure remains below typically 1$~$Torr, the gas is said to be in the molecular regime: the mean-free-path is long and the gas damping on the NEMS device has to be described by molecular shocks transferring momentum \cite{PREpaper,Yamamoto}. At higher pressure, the fluid can be described by the Navier-Stokes equations \cite{saderBeam}.
The cross-over between the two regimes is a complex issue that has attracted interest recently for both fundamental and practical reasons \cite{recentEkinci,bullard}. Besides, controlling thermal gradients in the cell and NEMS velocities, we believe that there is no relevant net (static or oscillatory) flow around the boundary in our experiments (see Supplementary Material \cite{SM}).

When the pressure is low enough such that the mean-free-path $\lambda$ is of the order of the gap $g$, we observe a reduction of the damping below the expected $\Delta f_{molec.} \propto P$ molecular law (Fig. \ref{figure2} inset). This was first reported in Ref. \cite{PRLus} for two other devices of different length $L$ and gap $g$. However at that time, the low-pressure analytic interpretation was not specific and a tentative power-law fit had been proposed. 
In the main graph of Fig. \ref{figure2} we plot the broadening normalized to the standard molecular law $\Delta f/\Delta f_{molec.}$ with respect to $\lambda/g$, the relevant Knudsen number in our problem. 
This data is compared to that of the two devices of Ref. \cite{PRLus}, analyzed in the same way. By construction all curves start at 1, and decrease for larger $\lambda/g$ with up to a factor of 10 reduction in damping, which is remarkable.

The shape of the measured curves in Fig. \ref{figure2} for different devices is rather similar; assuming that indeed the analytic dependences should be the same, we show that all data can be fit consistently by the same {\it Pad\'e approximant} leading to the same asymptotic laws:
\begin{equation}
\frac{\Delta f}{\Delta f_{molec.}} = \frac{1 + c \, \frac{\lambda}{g}}{1+ \left( c - \alpha \right) \frac{\lambda}{g} + \frac{c}{\alpha'} \left(\frac{\lambda}{g} \right)^{\!2} } , \label{fit}
\end{equation}
which gives $1+\alpha \frac{\lambda}{g}$ at first order ($P \approx 1~$Torr) and $\alpha' \frac{g}{\lambda}$ when the pressure is very low $P \ll 1~$Torr. The parameter $c$ then captures the rounded shape that joins these two limits in Fig. \ref{figure2} (see S.M. \cite{SM} for details).

\begin{figure}[h!]
\includegraphics[height=7.5 cm]{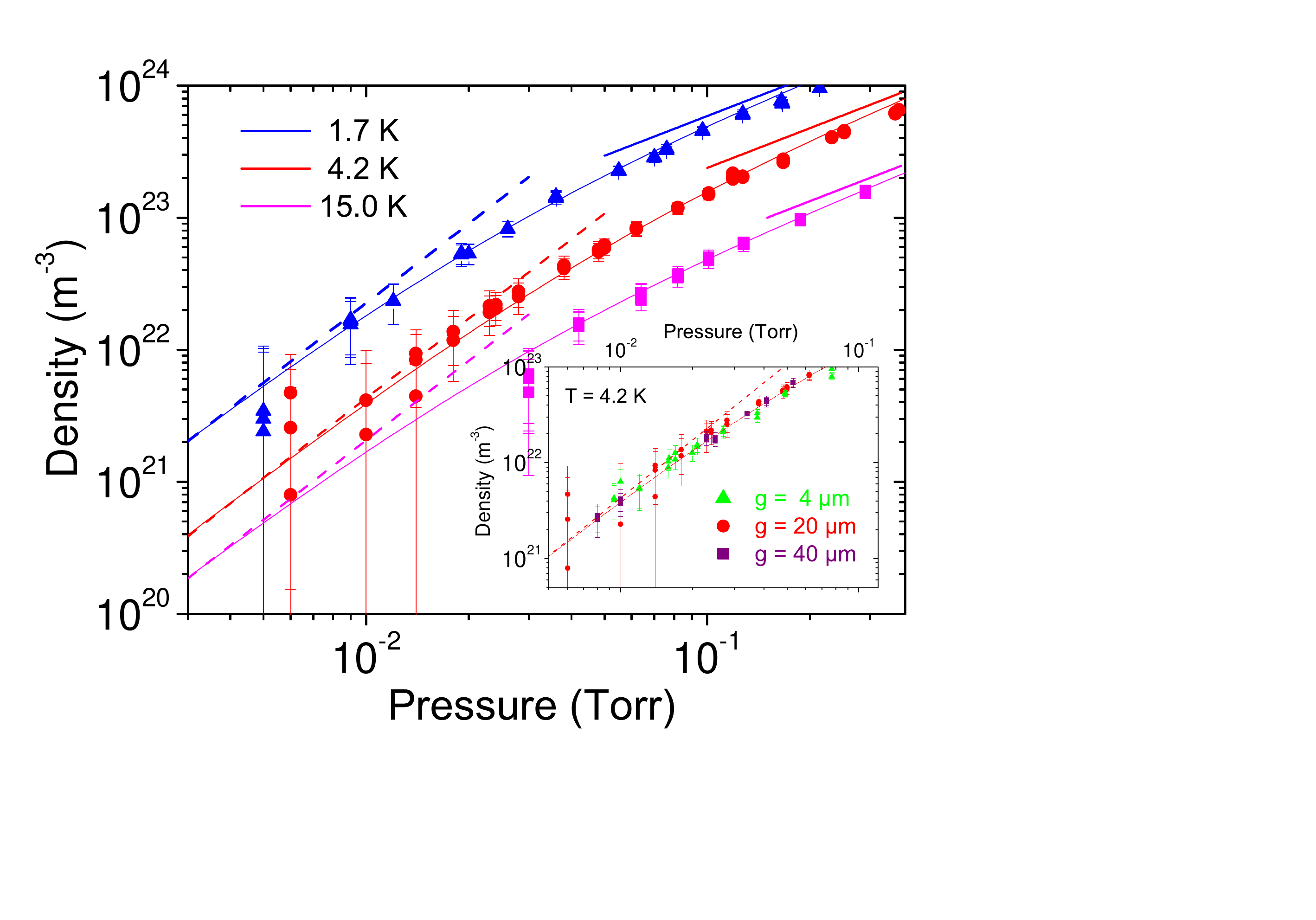}
\vspace{-1.5cm}\caption{\label{figure3} (Color online) Main: Effective density calculated at 3 different temperatures for the 150$~\mu$m long device. Thin lines correspond to fits obtained from Eq. (\ref{fit}) while the thick lines at high pressure ($\propto P$) and the dashed lines at low pressures ($\propto P^2$) are the asymptotes. Inset: Same data at 4.2$~$K for the 3 different devices, showing the independence towards $g$. }
\end{figure}

Remarkably, the parameter $\alpha$ is independent of the gap $g$, which proves that indeed the cross-over from the standard molecular regime to the boundary layer regime occurs when $ \lambda  \approx g $. However, $\alpha'$ is inversely proportional to $g$ meaning that when the NEMS device is deeply immersed into the boundary layer, the measured gas damping is {\it independent} of $g$ (Fig. \ref{figure3} inset). 
This is also to be expected, since in this limit $ \lambda  \gg g $ and $g$ cannot be a relevant lengthscale anymore: essentially the local probe senses molecular shocks {\it almost on} the boundary surface.

\vspace{1mm}

The first order deviation $\alpha \frac{\lambda}{g}$ to molecular scattering can be accounted for by adapting known kinetic theory models. The idea is that the boundary scattering on the surface induces deviations from the Maxwell-Boltzmann (MB) equilibrium distribution that propagate within the gas over a lengthscale commensurate with the mean-free-path. Furthermore, we shall demonstrate that the mathematical development interprets the measurements as a {\it rarefaction phenomenon} occurring within the boundary layer: locally, because of the deviations to MB the gas density is reduced. 

The starting point consists in noting that the NEMS probe is essentially a non-invasive sensor since $w \ll \lambda$ and  $w \ll g$ \cite{PRLus}. As such, all deviations to standard molecular damping have to proceed from the boundary scattering. On the wall, the collision integral $Q_{wall}(f,f)$ writes formally:
\begin{equation}
Q_{wall}(f,f) = \int \!\! \int \left( f' f_1'-f f_1\right) B\left[\Omega,\vec{v},\vec{v}_1\right] d \Omega \, d \vec{v}_1, \label{Qintegral}
\end{equation}
with $B\left(\Omega,\vec{v},\vec{v}_1 \right)$ the scattering kernel produced by the actual interaction on the surface, the notations corresponding to the process $\left\{ \vec{v}, \vec{v}_1 \right\} \rightarrow \left\{ \vec{v}', \vec{v}_1' \right\}$ \cite{Carlo}. The distribution function $f(\vec{v})$ verifies the Boltzmann equation; but on the wall there is {\it no reason} for the complex interactions between gas particles and adsorbed atoms to zero the collision integral, leading to $Q_{wall}(f,f)\neq 0$.


This implies \cite{Carlo} that $f \neq f_0$ (the MB distribution) close to the surface, and we assume the deviation to be small and regular enough to be expanded in powers of the particles' velocity field $\vec{v}$:
\begin{equation}
f(\vec{v}) = f_0 \left( 1+ Pl\left[\vec{v} \right] \right) , \label{ffunc}
\end{equation}
with $Pl\left(\vec{v} \right)$ a polynomial. This is essentially the approach first proposed by Grad \cite{Carlo,Chapman,grad,patterson,Sone}, but we do not assume here any particular polynomial form since we do not know the symmetries of the scattering kernel $B$. 
We proceed by applying the Chapman-Enskog method to the coefficients of the polynomial $Pl$ themselves \cite{Carlo,Chapman}: 
we assume that {\it each of them} can be developed in a series of $\frac{\lambda}{x}$. 



The combination of the two mathematical techniques (Grad and Chapman-Enskog) does not aim at calculating these parameters; it is essentially a {\it phenomenological macroscopic approach} which enables us to justify the analytic form Eq. (\ref{fit}) in its ''high pressure'' asymptotic dependence (the $P \approx 1~$Torr range, when $\frac{\lambda}{g} \approx 1$). The interesting aspect of the mathematical treatment lies in the fact that we do not need to stipulate the scattering kernel $B$. 
Furthermore, it is not a trivial expansion: the calculation has been performed up to order 4 in velocity, and order 3 in Knudsen number in order to demonstrate that indeed the approach is self-consistent. As a result, one obtains the macroscopic thermodynamical parameters temperature $T(x)=T_0+ \delta T(x)$, density $n(x)=n_0+ \delta n(x)$ and kinetic pressure along the $\vec{x}$ axis $P_{xx}(x) =P_0+ \delta P (x)$ as Taylor series in $\frac{\lambda}{x}$ depending on the near-field parameters introduced by the expansions. 

The key result is that the kinetic pressure $P_{xx}(x)$ is constant within the boundary layer ($\delta P [x]=0$), the temperature contains a second order correction at lowest order $\delta T(x) \propto \left( \frac{\lambda}{x}\right)^2$ leading to the known temperature jump phenomenon on the surface \cite{siewert}, while the density deviation is first order $\delta n(x) \propto  \frac{\lambda}{x}$. This is schematized in Fig. \ref{figure1} a), with the device probing the $x=g$ position in space: in this sense, the measured first order {\it decrease} of the friction force with respect to molecular damping {\it is due to the rarefaction of the gas in the boundary layer}. 
Details on the calculation can be found in Supplementary Material \cite{SM}.

This macroscopic approach is robust, provided all the expansions are defined. These hypotheses essentially mean that we consider the mathematical treatment far enough from the surface ($x=0$ is indeed pathological). What happens very close to the surface is an extremely complex problem as far as mathematics are concerned, far beyond the phenomenological approach. Note also that in the literature, the deviations from MB close to a wall are discussed usually in the framework of gas flows \cite{Carlo,patterson,dadzie,sader,saderII}; here, the complex nature of the interaction with the surface is the only source of deviations. The gas is at equilibrium, with a {\it continuous dynamic exchange} of atoms between the adsorbed ones and the gas boundary, but the distribution of velocities is  non-MB: the spatial gradients can be seen as due to a near-field force which originates on the boundary, from the presence of the adsobed atoms.

\begin{figure}[h!]
\includegraphics[height=8. cm]{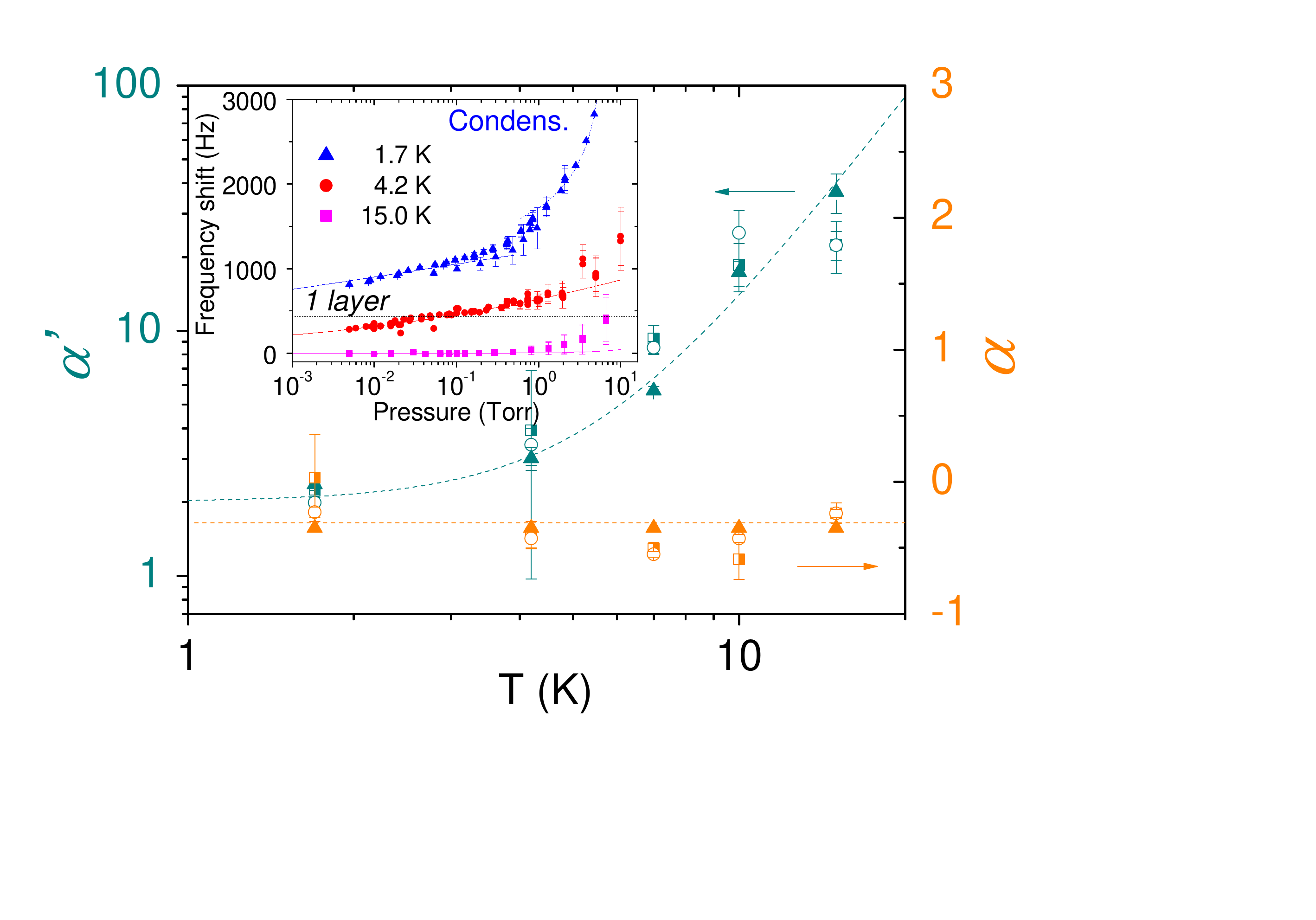}
\vspace{-1.5cm}\caption{\label{figure4} (Color online) Main: Leading-order coefficients extracted from the fits of Fig. \ref{figure2}. Different symbols stand for different fitting routines (see S.M. \cite{SM}). Dashed lines are guides for the eye. Inset: Measured frequency shifts due to adsorbed layers onto the NEMS surface for 3 temperatures. 
For 1.7$~$K and 4.2$~$K, data obtained on the different devices have been normalized to the $L=150~\mu$m one \cite{SM}.
At 1.7$~$K, the rapid growth when $P$ is increased corresponds to the condensation of a thin superfluid film \cite{SM,Dash,Pobell}, while at higher temperatures the weaker growth is a signature of the cross-over towards the Navier-Stokes regime. Lines at low pressures are fits based on Dubinin-Astakhov isotherms \cite{SM,DA,Rudzinski,Cerofolini,Hutson}, while the horizontal dashes correspond to a {\it dense} adsorbed monolayer \cite{Schildberg}.}
\end{figure}


By changing temperature $T$ we can also tune the mean-free-path $\lambda$ \cite{SM}. Besides, since the decrease in damping is essentially due to a decrease in density, we can compute an effective density $n_{ef\!f.} \equiv n$ in the boundary layer from the molecular damping expression $\Delta f_{molec.} \propto n$. We present in Fig. \ref{figure3} these measurements with the $L=150~\mu$m NEMS device at 3 different temperatures. 
We see that the data can again be very well fit by the {\it Pad\'e Approximant} Eq. (\ref{fit}). At high pressures, we find as we should the asymptotic deviation $\alpha \frac{\lambda}{g}$ for the effective density. Remarkably, the fit coefficient $\alpha$ appears to be also temperature independent; this is shown in Fig. \ref{figure4}.

On the other hand, at low pressures the effective density scales as $P^2$ (dashed lines in Fig. \ref{figure3}). This corresponds to the $\alpha' \frac{g}{\lambda}$ asymptotic behavior already discussed when we introduced Eq. (\ref{fit}). In the inset of Fig. \ref{figure3} we show that
the measured effective density does not depend on the gap $g$, as it should. The fit parameter $\alpha'$ is presented in Fig. \ref{figure4}; as opposed to $\alpha$, it {\it strongly depends on} temperature.

The fact that $\alpha$ is temperature-independent suggests that the first order deviation is essentially driven by the physical mismatch that the surface introduces in the problem, regardless of the {\it excitations} that it can support. 
In this sense, the first order deviation in Knudsen number $\lambda/g$ seems to be {\it universal}.
However in the other limit, $\alpha'$ depends on temperature, and seems to increase rather quickly when the temperature becomes equivalent to the {\it adsorption energy} (Fig. \ref{figure4}). In the inset of Fig. \ref{figure4}, we show how the NEMS resonance frequency shifts when mass is added through adsorbed layers. Even if the NEMS surface is physically not the same as the probed boundary (the spongy background in Fig. \ref{figure1} b), this gives us important information about the surface coatings present in the experimental cell \cite{SM};  in particular, we can estimate the number of adsorbed atomic layers.
The growth of $\alpha'$ correlated with the adsorption temperature proves that the surface coating plays an important role in the scattering mechanisms at very low pressures. Besides, it explains why the rarefaction effect could not be demonstrated at room temperature \cite{nanoEkinci}; indeed in Fig. \ref{figure3}, as the temperature is increased the cusp in the measured friction (change from $P^2$ to $P$ laws) becomes less and less visible.


In conclusion, we measured the friction force exerted by $^4$He gas at cryogenic temperatures on a NEMS device. 
When the pressure is very low, we report on a decrease from the standard molecular damping $\Delta f_{molec.} \propto P$. We explain how this effect can be interpreted in terms of a {\it rarefaction} of the gas near the surface boundary, induced by a deviation from the standard Maxwell-Boltzmann equilibrium distribution. 
This phenomenon can be seen as a near-field force propagating within the gas over a length commensurate with the mean-free-path $\lambda$, the Knudsen layer. Deep in the boundary layer, the effective density of the gas seems to scale as $P^2$ instead of $P$. All the experimental data can be fit using a simple {\it Pad\'e approximant} expression, demonstrating that the first order deviation from molecular damping is temperature and geometry independent. On the other hand, the $P^2$ dependence is also a function of temperature, strongly marked by the adsorption energy of $^4$He atoms on the chip surface. This demonstrates the importance of the dynamics of adsorbed atoms in this effect, and explains why room temperature experiments fail to reproduce it. The phenomenon should be accompanied by a temperature jump, and is clearly calling for further theoretical and experimental developments.

We thank J.-F. Motte, S. Dufresnes and T. Crozes from facility Nanofab for help in the device fabrication. We acknowledge support from the ANR grant MajoranaPRO No. ANR-13-BS04-0009-01 and the ERC CoG grant ULT-NEMS No. 647917. This work has been performed in the framework of the European Microkelvin Platform (EMP) collaboration.


\newpage
$\,$
$\,$

\begin{center}

{\huge Supplementary Information for \\}
{\huge Surface-induced near-field scaling in the Knudsen layer of a rarefied gas}
\end{center}

\section*{Device, magnetomotive scheme and setup}
\label{expesetup}

The device we used is shown in Fig. 1 b). Its characteristics are described in the main Article. It is made of high-stress silicon nitride (100$~$nm, with 0.9$~$GPa) on top of silicon. The device is patterned using e-beam lithography, and an aluminum evaporation (30$~$nm). The nitride is etched by Reactive Ion Etching (SF$_6$ plasma) using the aluminum as mask, and finally the underlying silicon is removed using XeF$_2$ dry etching. In order to obtain the deep trenches, a large undercut is also present on the structures. The spongy nature of the bottom is due to the etching. We compare and re-analyze data from two other (similar) devices described in Ref. \cite{PRLus}.

\begin{figure}[h!]
\includegraphics[height=7 cm]{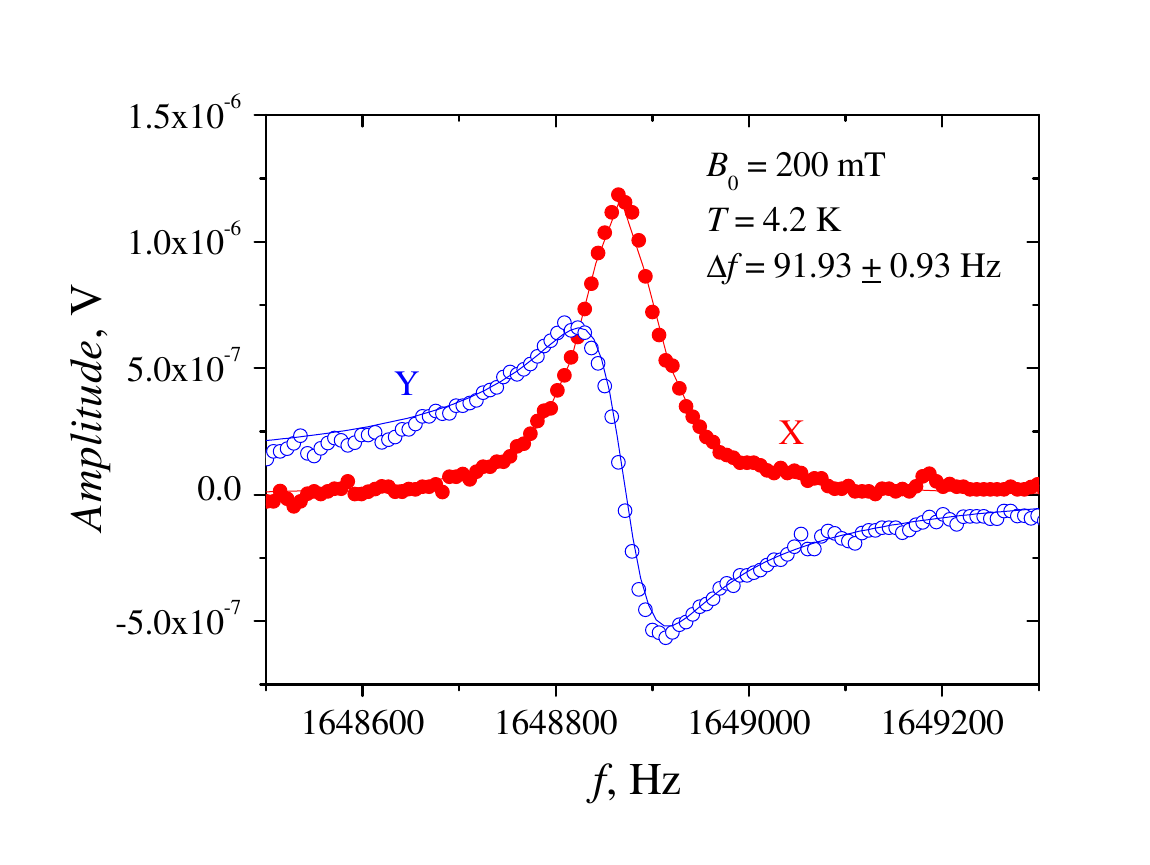}
\vspace{-0.3cm}\caption{\label{resone} Resonance line at 4.2$~$K in vacuum for a small field, in the linear regime. The fit is a lorentzian lineshape, with the two quadratures $X$, $Y$ obtained from the lock-in.}
\end{figure}

Actuation and detection are done with the magnetomotive scheme \cite{RSIus,SensActCleland}.
We drive the device with a current fed to the metallic layer via a 1$~$k$\Omega$ bias resistor $I(t)=I_0 \cos(\omega t)$.
The NEMS is glued to a copper sample holder, held in a cell connected to room temperature through a pumping/feeding line. Around the cell we have a coil producing a field $B_0$ orthogonal to the beam, and within the chip plane.
There is thus an out-of-plane Laplace force $F^0=\zeta_0 L B_0 I_0 \cos(\omega t)$ acting onto the device, with $L$ the length of the beam. When $\omega$ is close to $\omega_0$ (the resonance frequency of the first out-of-plane flexure), the beam moves (with motion amplitude in the center $x_0[t]$) and cuts the field lines, thus inducing a voltage $V_0 = -\zeta_0 L B_0 \, v_0(t)$, with $v_0(t) = \dot{x}_0(t)$ the velocity.
The parameter $\zeta_0$ is a mode-dependent number that can be easily calculated from the mode shape.  
We detect $V_0$ using a lock-in amplifier locked at the drive frequency $\omega$. A typical resonance line is shown in Fig. \ref{resone}.

\begin{figure}[h!]
\includegraphics[height=7 cm]{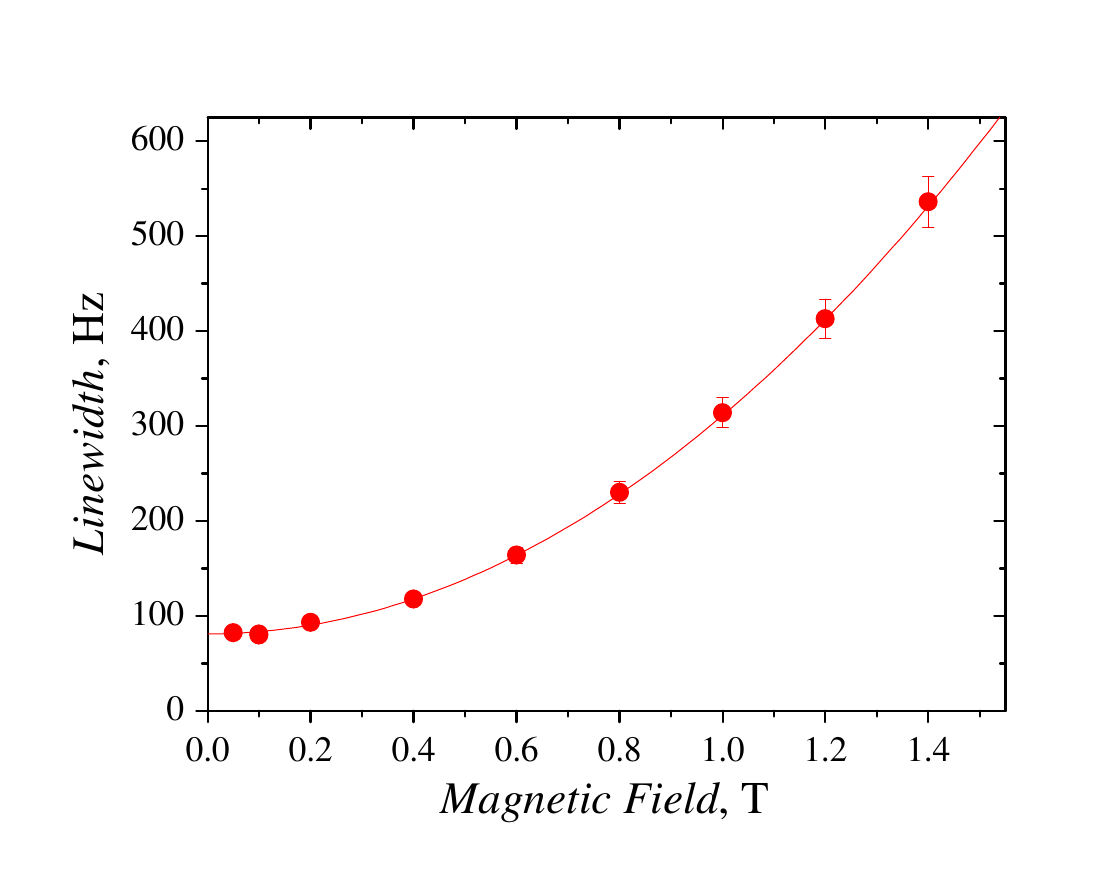}
\vspace{-0.3cm}\caption{\label{load} Resonance linewidth as a function of d.c. magnetic field, in vacuum at 4.2$~$K and linear regime.}
\end{figure}

In the magnetomotive scheme, the device is loaded by the impedance of the environment, leading to an extra damping $\propto B_0^2$ \cite{RSIus,SensActCleland}. We show this extra damping in Fig. \ref{load} with a fit. The measurements were done at low fields to limit the intrinsic damping that had to be subtracted, typically $B_0 \approx 200~$mT.

When the driving force is too large, the mechanical resonance becomes nonlinear with a pronounced Duffing shape. We kept here the drive low enough to be always Lorentzian. Besides, the gas damping itself seen by the NEMS can become nonlinear if the velocity $v_0(t)$ is too large. We checked experimentally that we were not in this situation; calculations can also be found in the Supplementary Material of Ref. \cite{PRLus}.
In particular, any effects arising from the finite size of the NEMS, which we treat here as a perfect local probe, would be velocity-dependent [see Eq. (\ref{molec}) below]. 

The pressure was monitored at room temperature from a Baratron\copyright $~$pressure gauge. We can calculate that pressure gradients along the pipe, generated by the temperature gradient (thermomolecular correction) are to be very small (Supplementary Material of Ref. \cite{PRLus}). We also estimated acoustic damping for our structures, which is also marginal. 
Besides, the experiment of 2014 \cite{PRLus} and today's results {\it were not obtained on the same cryostats}, and the two experimental cells were very different. Today's cell is much wider (4$~$cm diameter), and the cold finger is a thin copper plate (while in the old setup it was a massive copper rod). The pumping line is also much wider today (at least 1/2$~$cm all way long).
We re-measured, on the very same $L=300~\mu$m sample (which was still alive) of Ref. \cite{PRLus} the 4.2$~$K data already published: the curves are {\it rigorously the same}, the effect is perfectly genuine. Since in Ref. \cite{PRLus} the decrease of damping was shown to be independent of frequency (by testing different modes of the same structure), we could conclude as well that it was due to a finite-size effect, and not a finite-response-time effect \cite{recentEkinci}.

\begin{figure}[h!]
\includegraphics[height=7 cm]{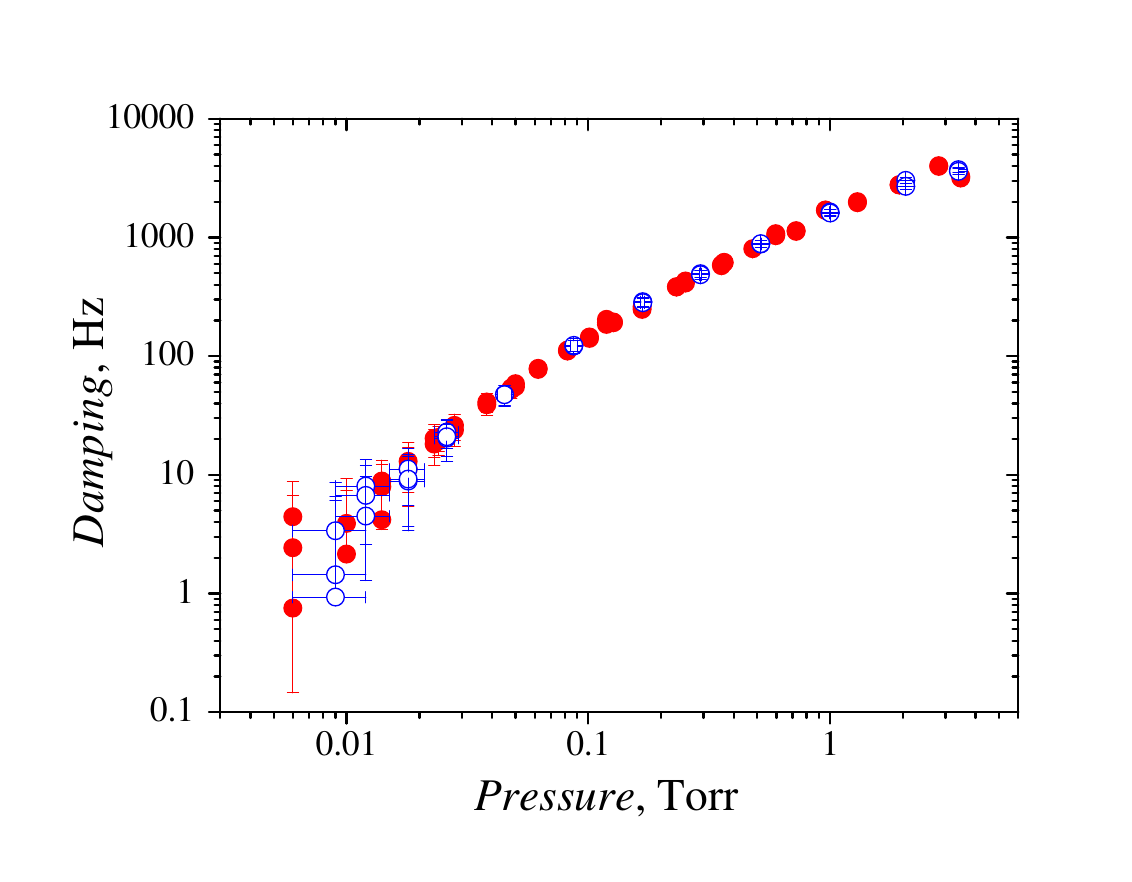}
\vspace{-0.3cm}\caption{\label{tempeffect} Gas damping at 4.2$~$K measured directly on our 150 $\mu$m device (full symbols) or with the regulation on from 1.4$~$K with about 100$~$mW heating (empty symbols).}
\end{figure}

The temperature of the sample was regulated with a heater mounted on the copper plate, and a thermometer measuring $T$ on the other end. Powers up to a few $100~$mW were required to maintain the chip at 15$~$K while the outside of the cell was at 4.2$~$K. 
In order to cool down below 4.2$~$K, the bath of the helium-4 cryostat was pumped down, with lowest reachable temperature of the order of 1.3$~$K.
Obviously, there is a temperature gradient within the gas in the cell: we estimate that in the worse situation, it leads to a thermal decoupling of $\delta T/T \approx 10 \times 20.\,10^{-6}/0.02/15 \approx 0.0007$ at the level of the NEMS. This is completely negligible, but to confirm it experimentally, we measured the gas damping curve obtained at the same temperature of 4.2$~$K by two means: either without any temperature regulation, or with the cryostat cooled at 1.4$~$K and the regulation on. The two curves are shown in Fig. \ref{tempeffect}, and do not display any difference. The pressure has thus to be essentially homogeneous in this setup, even though the gas density is not; which is the same argument as for the absence of thermomolecular corrections between 4$~$K - 300$~$K.
Finally, the adsorption isotherms in Fig. \ref{adsorbfig} are fit for $P, T$ with a {\it single} set of parameters consistently. If gradients were present, this analysis would not match.

\section*{Pad\'e approximant fits}
\label{Pade}

All our data of molecular friction could be fit by $\Delta f_{gas} = \Delta f_{molec.} \, Pad\acute{e}[\lambda/g]$ with $Pad\acute{e}[\lambda/g]$ given in Eq. (1). The standard molecular damping expression $\Delta f_{molec.}$ is described in Supplementary Material of Ref. \cite{PRLus}. It writes:
\begin{equation}
\Delta f_{molec.} =\frac{ 2 K \rho_g \bar{v}_g  w \, \xi_0 }{m_{nems}}, \label{molec}
\end{equation} 
with $\bar{v}_g$ the average velocity $\propto \sqrt{T}$ (not the root-mean-square; an equivalent expression can be written with it), $\rho_g $
the gas mass density $\propto P/T$, $w$ the width of the NEMS probe (300$~$nm), and $\xi_0 =\int \Psi_0(z)^2 dz$ a mode-parameter (the same as the one of the mode mass), with $\Psi_0(z)$ the fundamental flexure mode shape. Since $\xi_0$ appears also in the mode mass $m_{nems}$, the damping $\Delta f_{molec.}$ is independent of the mode, which had been verified in Ref. \cite{PRLus}. Therefore the molecular damping scales as $P/\sqrt{T}$, see Fig. 3 asymptotes (thick lines at high pressure).

In Eq. (\ref{molec}) the numerical prefactor $K$ close to 1 depends on the reflection mechanisms on the NEMS surface. A simple modeling including a specular fraction $p$ leads to $K= p(1+o[v_0]) + (1-p) (1+\frac{\pi}{4} \sqrt{\frac{T_{NEMS}}{T}}+o[v_0])/2$ (S.M. Ref. \cite{PRLus}).
The term $o[v_0]$ is a small correction that could be created on the NEMS surface by its own velocity field profile \cite{patterson}; it is found to be negligible. Depending on $p$ the $K$ parameter is found to deviate from 1 by not more than about 10$~$\%, if $T_{NEMS} \approx T$. We therefore simply fit it on data, wich amounts at choosing roughly $p \approx 0.5$ \cite{PRLus}.

\begin{figure}[h!]
\includegraphics[height=7 cm]{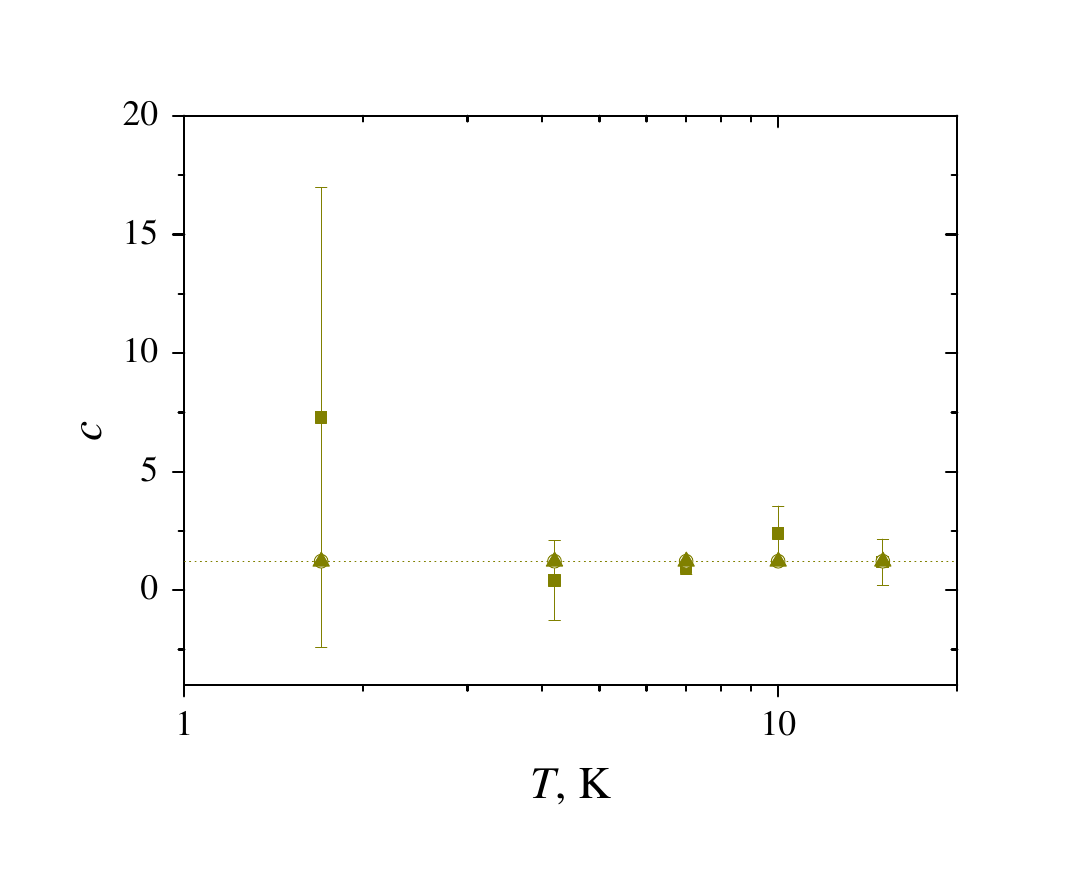}
\vspace{-0.3cm}\caption{\label{cparam} Fit parameter $c$ as a function of temperature for our $L=150~\mu$m sample. The symbols have the same meaning as in Fig. 4 (see text).}
\end{figure}

In Fig. 4 we present the Pad\'e approximant coefficients $\alpha$ and $\alpha'$ fit on the data. In Fig. \ref{cparam} we also show the parameter $c$, which captures the rounded part of the curves. 
In order to estimate the robustness of the fits, and since some coefficients ($\alpha$ and $c$) seemed to be temperature-independent, we tried different fitting routines differentiated by the symbols used. First, we left all coefficients free (marble squares). Then, we fixed $c$ constant, and let the computer fit the two other coefficients (empty circles). Finally, we fixed both $c$ and $\alpha$ and let $\alpha'$ be fit (full triangles). As a result, all fits are in very good agreement, and we seem to conclude that $\alpha \approx -0.3$ for all temperatures. 

\section*{Adsorbed layers on NEMS}
\label{adsorbed}

The measurements of $^{4}$He adsorption isotherms by our NEMS were first reported in Supplementary Material of Ref. \cite{PRLus}. Here we do a more systematic isotherm study as a function of $T$. The formation of adsorbed layers leads to an increased mass of the NEMS and manifests itself by a corresponding resonance frequency shift:
\begin{equation}
\delta f=f_{0}\sqrt{\frac{m_{nems}}{m_{nems}+m_{ads}}}-f_{0} \approx -f_{0}\frac{1}{2}\frac{m_{ads}}{m_{nems}},
\end{equation}
where $m_{nems}$ is the mass of NEMS and $m_{ads}$ is the added mass due to adsorbed layers.

\begin{figure}[htt]
\includegraphics[height=6.5 cm]{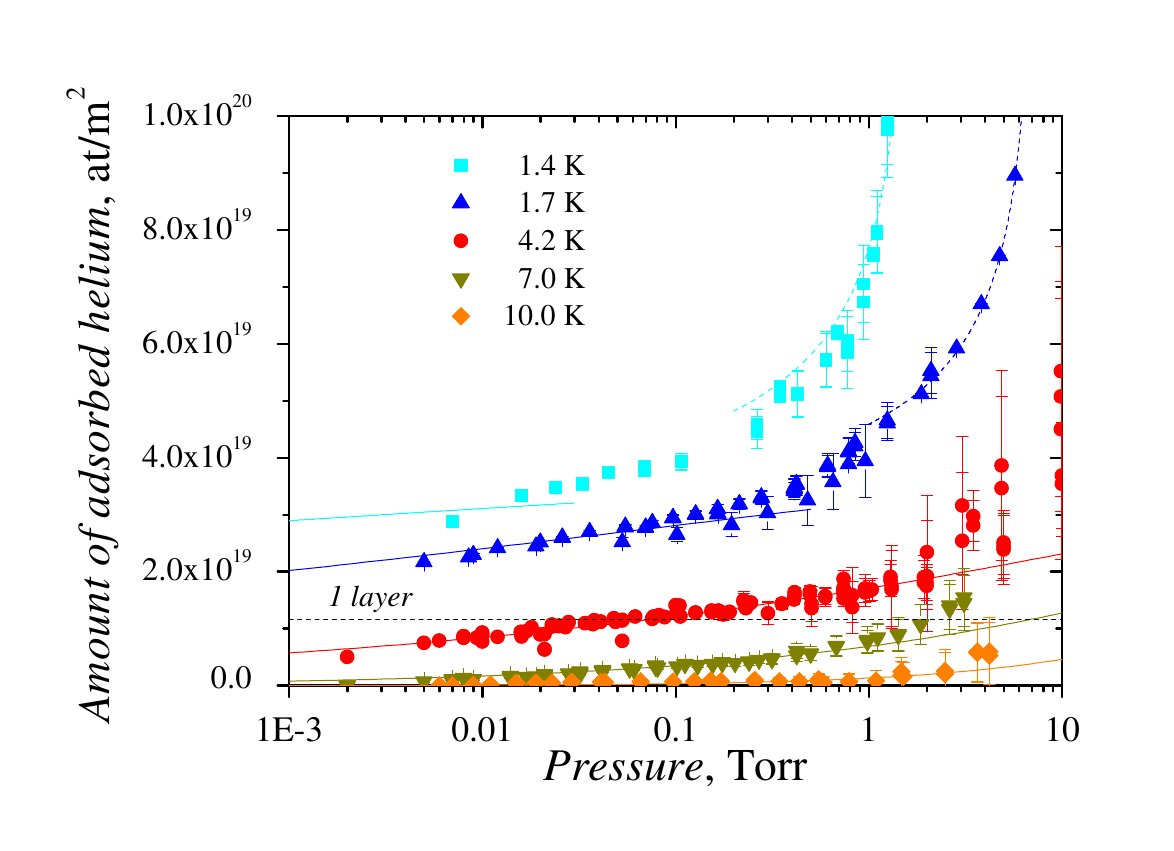}
\vspace{-0.3cm}\caption{\label{adsorbfig}Adsorption isotherms for three different devices (100, 150 and 300$~\mu$m long, scaled on 150$~\mu$m data) at 4.2$~$K, and data for the 150$~\mu$m device at different temperatures. The solid lines are fits by "heterogeneous" Dubinin-Astakhov (D-A) isotherm, Eq. (\ref{adsorb}). The dashed lines are fits by adsorption isotherms for film growth, Eq. (\ref{film}). Same fits as Fig. 4 inset.}
\end{figure}

By measuring the frequency shift, we can then recompute the adsorbed mass $m_{ads}$ as a function of pressure $P$ at different temperatures $T$. We express this mass in units of adsorbed helium density (Fig. \ref{adsorbfig}), using the bare atom mass $m_{He}$ = 6.65 10$^{-27}$ kg and the geometrical surface of the NEMS.
We can convert it into an equivalent number of adsorbed layers using a layer density $\rho_{ads}$ = 11.6 10$^{18}$ at/m$^{2}$ (see for example Ref. \cite{Schildberg}).
This density corresponds to the so-called incommensurate solid, the dense first layer of adsorbed $^4$He.

The surface of the NEMS is quite disordered at the scale of atoms (the Al grains are about 20 nm). In this case the mechanism of adsorption is somehow similar to that of pore-filling rather than layer-by-layer surface completion and the adsorption isotherms can be described by the Dubinin-Astakhov (D-A) equation \cite{DA}:
\begin{equation}
N=N_{m} \exp\left \{ -\left ( \frac{k_B T \ln \left[\frac{P_{s}}{P}\right] }{E} \right )^{\!n} \right \} , \label{adsorb}
\end{equation}
where $N_{m}$ represents the limiting amount of adsorption at saturated pressure $P_{s}$, $k_B$ is Boltzmann's constant and $E$ is the characteristic energy of adsorption. The term $n$ has been referred to in the literature as a "surface heterogeneity factor" \cite{Rudzinski}. We could fit our data on adsorption (solid lines in Fig. \ref{adsorbfig}) at low coverages by Eq. (\ref{adsorb}) with the following fixed parameters: $N_{m} \approx $ 3 layers, $E/k_B \approx30~$K and n $\approx$ 1.3. 

The critical point of $^4$He is $T_c=5.19~$K, $P_c=2.25~$bar. Above this temperature, the pressure $P_s$ is not defined anymore. However, Eq. (\ref{adsorb}) still describes very well the data, with only a logarithmic sensitivity to the actual value of $P_s$. We thus chose to fit with $P_s=P_c$ in this range (data at 7, 10 and 15 K), with no extra free parameter. 

\begin{figure}[htt]
\includegraphics[height=7 cm]{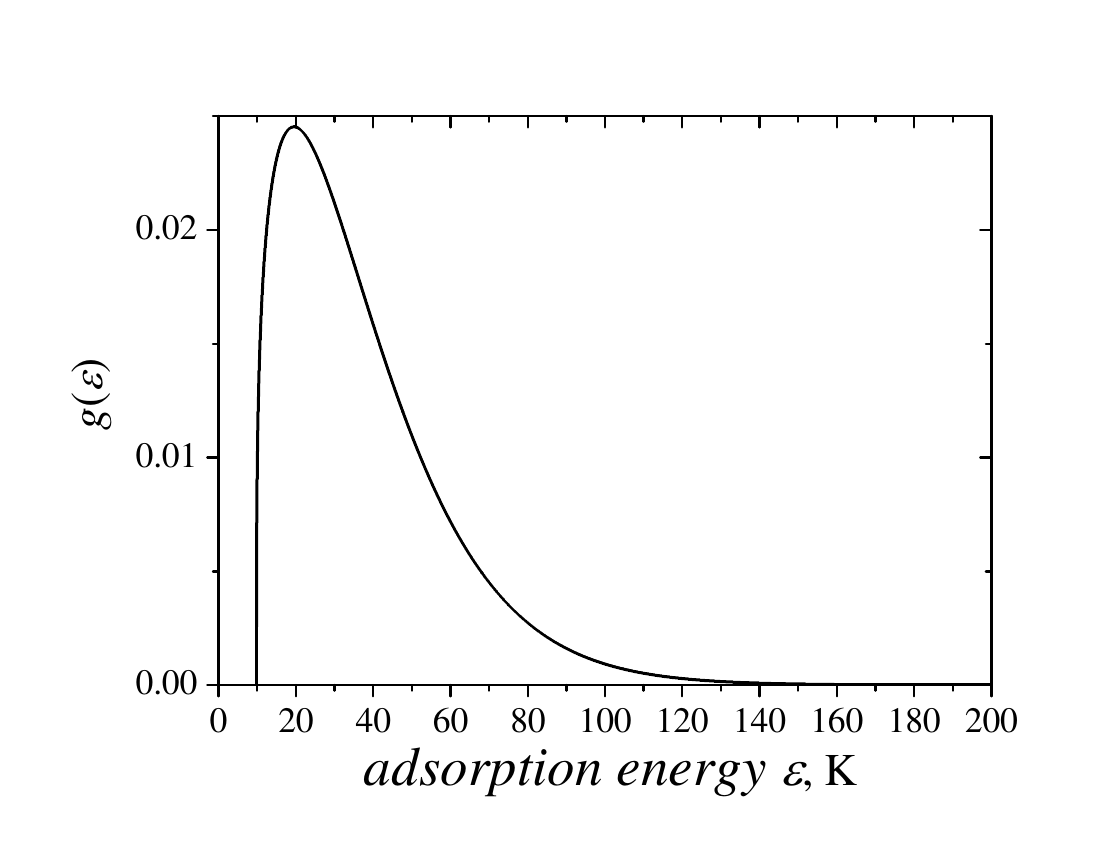}
\vspace{-0.3cm}\caption{\label{gdistrib}Adsorption energy distribution $g\left ( \varepsilon  \right )$ obtained for NEMS surface.}
\end{figure}

Eq. (\ref{adsorb}) stands for inhomogeneous adsorption surfaces. To find the adsorption energy distribution function it was assumed \cite{Cerofolini} that the heterogeneous surface consists of a family of non-interacting energetically homogeneous zones. If we consider as the effective local isotherm the Langmuir one and as the overall isotherm the D-A one, then we obtain the approximate energy distribution associated with the D-A equation, Eq. (\ref{adsorb}) \cite{Hutson}:
\begin{equation}
g\left ( \varepsilon  \right )=n\frac{\left ( \varepsilon -\varepsilon _{0} \right )^{n-1}}{E^{n}} \exp\left (- \left ( \frac{\varepsilon -\varepsilon _{0}}{E} \right )^{n} \right ),
\end{equation}
where $\varepsilon_{0}$ corresponds to the local minimum of adsorption potential resulting from intermolecular forces ($\sim$ 10 K for He atoms). From this adsorption energy distribution (plotted in Fig. \ref{gdistrib}), we found the adsorption potential to be maximum at about 20 K with a non-symmetic 30$~$K spread around the peak value.  

The adsorption process changes at high coverages when the effects of heterogeneity are healed out. When the layer thickness becomes comparable with the scale of lateral heterogeneity of the substrate, the topmost layers "see" an averaged value of binding energy and the thickness tends towards a uniform value \cite{Dash}. We could fit the data at high coverages ($N>N_{m} \approx$ 3 layers) by treating the solid as a uniform continuum where the interaction potential between a gas atom and the solid surface at a distance $d$ is of van der Waals type and given by:
\begin{equation}
u\left ( d \right )=-\frac{2\pi \varepsilon n_{s}\sigma ^{6}}{3d^{3}}=-\frac{\alpha}{d^{3}},
\end{equation}
where $n_{s}$ is the atomic density of the solid (60 at/nm$^{3}$ for Al, see refs. in Ref. \cite{Pobell}). Taking into account that for $^4$He atoms $\varepsilon / k_{B}$ = 10.22 K, $\sigma$ = 0.256 nm and 1 layer corresponds to 0.36 nm thickness, we get $\alpha$ = $\alpha_{K}N_{m}^{3}$ = 7.73 (K layers$^{3}$). 

Two cases can be described. The first one is for $P>P_s$, liquid is present in the cell (this is {\it not} our case). If $\mu_{layer}$ is the chemical potential of the adsorbed layer and $\mu_{0}$ is the one of the bulk liquid, then we have \cite{Pobell}:
\begin{equation}
\mu_{layer} = \mu_{0}+mgh-\frac{\alpha}{d^{3}},
\end{equation}
where $mgh$ is the gravitational term. In thermal equilibrium an atom has to have the same chemical potential on the surface of the bulk liquid and on the surface of the layer; this condition leads to:
\begin{equation}
d=\left(\frac{\alpha}{mgh}\right)^{1/3}.
\end{equation}  
If there is gas but no bulk liquid (as in Fig. \ref{adsorbfig}, with $P<P_s$) we have an "unsaturated film", whose thickness depends on the gas pressure $P$ which is:
\begin{equation}
P(h)=P_{s} \exp \left(-\frac{mgh}{k_{B}T}\right).
\end{equation}
Therefore the layer thickness in terms of adsorbed layer density is:
\begin{equation}
N=N_{m} \left(\frac{-\alpha_{K}}{T \ln \left[\frac{P}{P_{s}}\right]} \right)^{1/3}, \label{film}
\end{equation}
where $N_{m}$ is obtained from fitting the data at low coverages [using Eq. (\ref{adsorb})]. The fit of the data at high coverages is presented in Fig. \ref{adsorbfig} by dashed lines with no free parameters. These growth of layer thickness near the $^{4}$He saturated vapour pressure is associated with the formation of a superfluid $^{4}$He film \cite{Pobell}, although in our experiments we do not probe any superfluid properties of this film.

\section*{Grad-Chapman-Enskog model}
\label{maths}

The analytic fit produced at intermediate pressures (when $\lambda/g \approx 1$) can be justified, to some extent, from theoretical arguments. The Taylor expansion of $Pad\acute{e}[\lambda/g]$ is a polynomial that originates from some function of $\lambda/x$, which is assumed to be regular enough to be expanded, and evaluated at $x=g$ where the NEMS is. Of course when $x \rightarrow 0$ this becomes completely invalid.

We write the Maxwell-Boltzmann (MB) distribution as:
\begin{eqnarray}
f_0(\vec{v})  & = & \\ \nonumber
& &\!\!\!\!\!\!\!\!\!\!\!\!\!\!\!\!\!\!\!\!\!\!\!\!\!\!\! \left( \frac{\beta}{\pi}\right)^{3/2} \!\!\!\exp \left[ - \beta \left( [v_x-v_{x0}]^2 + [v_y-v_{y0}]^2 + [v_z-v_{z0}]^2 \right) \right],
\end{eqnarray}
with $\beta=m_{H \!e}/(2 k_B T)$, where $m_{H \!e}$ is the mass of a $^4$He atom and $k_B$ Boltzmann's constant. For the sake of being as generic as possible, we keep for the time being in the expression a {\it macroscopic flow} $\{v_{x0},v_{y0},v_{z0}\}$. 

As discussed in the core of the Article, in the boundary layer the MB distribution {\it is not} the equilibrium distribution for the gas particles. 
However, the wall and the bulk of the gas are in equilibrium at the same temperature $T_0$ by definition, and we describe a situation were we do not impose any macroscopic flow; if a {\it local} flow $\{v_{x0},v_{y0},v_{z0}\}$ exists, it would be induced by the surface scattering itself. Furthermore, we shall describe here the behavior of the gas in a region which is not too close from the surface boundary; thus, the corrections to the usual MB distribution shall be small, and the distribution $f$ that is sought is most certainly a smooth function, with a single peak structure. It is then natural to follow the development first introduced by Grad in the framework of the moment theory \cite{Carlo,Chapman,grad,patterson}. We develop the equilibrium distribution around the MB one in a power series of the velocity, assuming that all coefficients are small and that the development converges. This leads to the definition Eq. (3).

Defining the volumic gas density as $n(\vec{x})$, the density of particles per unit volume and unit velocity in the phase space $\{\vec{x},\vec{v}\}$ writes:
\begin{equation}
n \,f = n(\vec{x}) f_0(\vec{v}) \left[1+ Pl(\vec{v})\right], \label{nf}
\end{equation}
with $ Pl(\vec{v}) $ the polynomial deviation to MB. We write it explicitly, developing at 4rth order (we display here only the 2 first orders in order to keep the length tight):
\begin{eqnarray}
Pl(\vec{v}) & = & a_1 \sqrt{\beta} \, (v_x-v_{x0})  + a_2 \sqrt{\beta}\, (v_y - v_{y0}) \\ \nonumber 
& + & a_3 \sqrt{\beta}\, (v_z - v_{z0}) \\ \nonumber
& + & a_{11} \beta \,(v_x-v_{x0})^2 +a_{22} \beta\, (v_y- v_{y0})^2 \\ \nonumber
& + & a_{33} \beta\, (v_z- v_{z0})^2  \\ \nonumber
& + & a_{12} \beta \, (v_x-v_{x0}) (v_y-v_{y0}) \\ \nonumber 
& + & a_{13} \beta \, (v_x-v_{x0}) (v_z-v_{z0}) \\ \nonumber 
& + & a_{23} \beta \,(v_y-v_{y0}) (v_z-v_{z0}) + \ldots \, .  
\end{eqnarray}
In the above equation, the development has been normalized to $\beta$ such that the $a_{ijk\ldots}$ are numbers with no dimensions.
Clearly, the positions of the indexes $ijk\ldots$ are equivalent. 
In order for $f$ to be normalized (i.e. $\left\langle f\right\rangle=1$, with $\left\langle \dots \right\rangle= \int \!\! \int \!\! \int \dots dv_x dv_y dv_z$), for the flow terms to be the actual averages of the velocities (e.g. $\left\langle v_x\right\rangle=v_{x0}$), the coefficients should verify some sum rules \cite{patterson}. Besides, the temperature $T$ is defined by $\frac{1}{2} m_{H \!e} \left\langle (v_x-v_{x0})^2+(v_y-v_{y0})^2+(v_z-v_{z0})^2 \right\rangle = \frac{3}{2} k_B T $ which also imposes some relations on the coefficients. The kinetic pressure terms $P_{a b}$ with $a,b=x,y,z$ are defined as the flow of momentum $m v_a$ through the surface normal to $\vec{b}$.
As such, the $P_{a b}$ with $a \neq b$ derive from the viscosity tensor while the thermodynamic pressure is defined as $P=\frac{1}{3} (P_{x x}+P_{y y}+P_{z z})$. It happens that $P = n k_B T$ is always valid, but $P_{x x}$ could deviate from this simple ideal gas law because of some of the $a_{ijk\ldots}$ coefficients.

In the $Pl(\vec{v})$ expression each $a_{ijk\ldots}(\vec{x})$ depends on position $\vec{x}$: it has to be zero in the bulk when $x \rightarrow +\infty$, and shall be nonzero but $\left| a_{ijk\ldots} \right| \ll 1$ close to the wall, for $x \neq 0$. The density $n \, f$, Eq. (\ref{nf}) verifies the Boltzmann equation:
\begin{equation}
\frac{\partial \left( n\,f\right)}{\partial t} + \vec{v}. \vec{\nabla}_{x} \left( n\,f \right) = Q(f,f) , \label{MBeq}
\end{equation}
with $Q(f,f)$ the collision integral and $\vec{\nabla}_{x}$ the spatial gradient operator. 

We search a steady-state solution, so no explicit time-dependence appears and the first term above is zero. Second, for symmetry reasons all space variables shall restrict to $x$ only, and no explicit $y, z$ appears either.
However, the velocity field remains defined in the 3 directions of space: the problem is truly 3D.
To further simplify the problem and make it tractable, we apply the Chapman-Enskog \cite{Carlo,Chapman} method to the $a_{ijk\ldots}$ coefficients themselves. We assume that if we are not too close to the wall, the corrections to the MB equilibrium function can be developed in power series of $\lambda/x$. We write at third order:
\begin{eqnarray}
a_{ijk\ldots} (x) & = & b^{(1)}_{ijk\ldots} \frac{\lambda}{x}  \\ \nonumber
& + & b^{(2)}_{ijk\ldots} \left( \frac{\lambda}{x}\right)^2+ b^{(3)}_{ijk\ldots} \left( \frac{\lambda}{x}\right)^3 + \ldots ,
\end{eqnarray}
with $b^{(n)}_{ijk\ldots}$ numbers with no dimensions. These numbers are characteristic of how the boundary scattering is seen, at a macroscopic level, by the gas for $x>0$ towards the bulk. The question is now to see how we can solve Eq. (\ref{MBeq}) in a self-consistent way in order to define $n(x)$, $T(x)$ and $P_{xx}(x)$. 

In order to explicitly compute these thermodynamic properties, we need to know $Q(f,f)$. It is defined as Eq. (2) in the main text explicitly at the boundary, as a function of the boundary scattering kernel $B\left(\Omega,\vec{v},\vec{v}_1 \right)$. For $x>0$ an identical expression holds with $B_{near\,field}\left(\Omega,\vec{v},\vec{v}_1 \right)$ the propagation of this kernel within the bulk of the gas. Evidently when $x\rightarrow +\infty$ we shall recover $Q(f,f)=0$ and the MB distribution at equilibrium.

The key is that we do not need here to specify either $B$ or $B_{near\,field}$; the problem can be treated self-consistently by realizing that the collision integral has to be as well expandable in velocities, and in a power series of  $\lambda/x$ in order for Eq. (\ref{MBeq}) to apply.
One writes:
\begin{eqnarray}
Q(f,f) &=& n_0 \frac{ \exp \left[ -\beta_0 (v_x - v_{x0})^2  \right]  \sqrt{\beta_0}}{\sqrt{\pi}} \times \label{kernel} \\ 
  &  & \!\!\!\!\!\!\!\!\!\!\!\!\!\!\!\!\!\!\!\!\!\!\!\!\!\!\!\!\!\!\!\!\!\!\!\frac{\left( \kappa_1 (v_x - v_{x0}) + \kappa_2 \sqrt{\beta_0} (v_x - v_{x0}) ^2 + \kappa_3 \beta_0 (v_x - v_{x0}) ^3 + \ldots \right)}{\lambda}  \nonumber 
\end{eqnarray}
with $\beta_0$ defined from $T_0$, and the local mean flow $v_{x0}$ still present in the expression. $n_0$ is the bulk equilibrium density. To match the same velocity order than the Grad-like 4rth order expansion, Eq. (\ref{kernel}) shall be written at order 5.
The $\kappa_i$ coefficients are functions of $\lambda/x$ such that:
\begin{eqnarray}
\kappa_i(x) &=& + \tau_{i,2} \left( \frac{\lambda}{x} \right)^2 + \tau_{i,3} \left( \frac{\lambda}{x} \right)^3 \label{kappai} \\ \nonumber
&  + &  \tau_{i,4} \left( \frac{\lambda}{x} \right)^4 + \ldots ,
\end{eqnarray}
written at fourth order, with the $\tau_{i,j}$ numbers to be defined. Note that because of the derivation in the left hand side of Eq. (\ref{MBeq}), there is no first order term above. 
Finally, the mathematical problem is fully set by defining $n(x)=n_0+\delta n(x)$, $T(x)=T_0+\delta T(x)$ and $P_{xx}(x)=P_0+\delta P(x)$ with the deviations expressed as power series of $\lambda/x$ as well, for the same reasons as for the other functions introduced. The values $n_0$, $P_0$ and $T_0$ are by definition the equilibrium values obtained in the bulk at $x \rightarrow + \infty$, and {\it imposed} by the wall at $x=0$ precisely. We write 
$\delta T(x)/T_0 = \alpha_{T_1} \frac{\lambda}{x} + \alpha_{T_2} \left( \frac{\lambda}{x}\right)^2 + \alpha_{T_3} \left( \frac{\lambda}{x}\right)^3+ \ldots$ explicitly.

Solving then the problem is straightforward but tedious; we use a Mathematica\copyright $~$ code to achieve this. 
However, in addition to the sum rules already quoted, many terms have to be 0. We obtain first that the local velocity flow {\it must} be zero ($v_{x0}=v_{y0}=v_{z0}=0$): there is no local macroscopic motion induced by the boundary. Second, the even terms in $Pl(\vec{v})$ with respect to $x$ shall also be zero, i.e. $a_1=a_{111}=a_{122}+a_{133}=0$ for a development truncated at 4th order. Note that many terms in the $a_{ijk\ldots}$ expansion are irrelevant, and eventually only $a_{11}$, $a_{1111}$ and $a_{1122}+a_{1133}$ matter here. Third, the even terms in the velocity development of $Q(f,f)$, Eq. (\ref{kernel}) are also all zero: $\kappa_2=\kappa_4=0$, when stopped at 5th order. Only $\kappa_1$, $\kappa_3$ and $\kappa_5$ matter.

It turns out that the $\tau_{5,j}$ are defined from the $\tau_{3,j}$ and $\tau_{1,j}$. The $\tau_{i,2}$ terms ($i=1,3$) are {\it linear combinations} of the relevant $b^{(1)}_{ijk\ldots}$ terms {\it and} the temperature coefficient $\alpha_{T_1}$, while $\tau_{i,3}$ are {\it quadratic combinations} of $b^{(1)}_{ijk\ldots}$, $\alpha_{T_1}$ {\it plus} linear ones of $b^{(2)}_{ijk\ldots}$, $\alpha_{T_2}$. Similarly for the $\tau_{i,4}$ coefficient: it is a cubic combination of the $b^{(1)}_{ijk\ldots}$, quadratic times linear in $b^{(1)}_{ijk\ldots}$, $b^{(2)}_{ijk\ldots}$ and linear in $b^{(3)}_{ijk\ldots}$ (associated again with the $\alpha_{T_i}$ of similar order). 

The problem is thus perfectly self-consistent and the collision integral $Q(f,f)$ is defined from the boundary scattering coefficients $b^{(n)}_{ijk\ldots}$. In addition to the initial convergence imposed to the series we introduced, Eq. (\ref{kappai}) should obviously also be convergent. If this is taken as granted, the problem is solved. We find that:
\begin{itemize}
\item There is no renormalization on the pressure $P_{xx}$, i.e. $\delta P=0$. The force per unit surface orthogonal to the wall is {\it constant} from the bulk to the wall. Even if the calculation has been done only to 4th order in $\lambda/x$, we believe that this is a {\it robust} property.
\item $n(x)$ has a first order correction: $\delta n(x)/n_0 = -\frac{1}{2} \left(2 \, b^{(1)}_{11} +6 \, b^{(1)}_{1111} + \left[ b^{(1)}_{1122}+b^{(1)}_{1133} \right] \right) \frac{\lambda}{x}$. This is indeed the rarefaction phenomenon, expressed at lowest order. The coefficients $b^{(n)}_{ijk\ldots}$ involved are only first order $n=1$.
\item The temperature varies only at second order $\delta T \propto \left( \frac{\lambda}{x} \right)^2$, i.e. $\alpha_{T_1}=0$. Furthermore, the $\alpha_{T_2}$ coefficient itself appears to be quadratic in the $b^{(n)}_{ijk\ldots}$ (as the $\tau_{i,3}$ described above). In this sense, the temperature gradient is much weaker (i.e. second order) than the density one and the phenomenon is mainly a density reduction.
\end{itemize}
The combination of orders in the resulting expressions seems to show that indeed, if the initial series are convergent, the solution ones also are. Furthermore, solving for the thermodynamic properties $n(x)$, $P_{xx}(x)$ and $T(x)$ happens to reduce to find the $\alpha_{T_i}$ coefficients of the $\delta T$ expansion. At the order discussed here, $\alpha_{T_2}$ is found to depend on $\alpha_{T_3}$ (and of course the  $b^{(n)}_{ijk\ldots}$ as already said), but $\alpha_{T_3}$ is undefined. We believe that this is also a {\it robust} property, and that an expansion to order $n+1$ would lead to expressions for all $n$ first coefficients, but the last $n+1$ one.

We can finally compute the friction force on the NEMS, taking $x=g$. Using the same writing as Eq. (\ref{molec}), it amounts to obtain the new prefactor $K_{near\,field}$ in the presence of non-MB deviations replacing the $K$ appearing in the molecular damping expression. The calculation leads to:
\begin{eqnarray}
K_{near\,field} & = & p + (1-p) \frac{1}{2} \left(1+\frac{\pi}{4} \sqrt{\frac{T_{NEMS}}{T_0}} \right) \\ \nonumber
&\!\!\!\!\!\!\!\!\! + & \!\!\!\!\!\! \frac{1}{2} \left(p+(1-p) \left[\frac{\pi}{4} \sqrt{\frac{T_{NEMS}}{T_0}}+ \frac{1}{2}\right] \right) \frac{\delta n(g)}{n_0} \\ \nonumber
&\!\!\!\!\!\!\!\!\! - & \!\!\!\!\!\!\frac{1}{8} (1+p) \, b^{(1)}_{1111} \frac{\lambda}{g},
\end{eqnarray}
at first order in $\lambda/g$, with $\delta n(g)/n_0$ limited to its first order as given above. We kept a NEMS temperature $T_{NEMS}$ potentially different from the bulk $T_0$, and a specular fraction $p$.
The first term above is the usual molecular damping term.
As a result, we see that indeed the dependence of the friction force to the boundary scattering terms $b^{(1)}_{ijk\ldots}$ is {\it almost the same} as the density $\delta n(x)$, i.e. the discrepancy is only in the last term, $-\frac{1}{8} (1+p) \, b^{(1)}_{1111} \frac{\lambda}{g}$. 
This again confirms that the measured deviation to standard molecular damping is indeed a rarefaction effect in the Knudsen layer.

The fit parameter $\alpha$ appears then as a rather involved combination of the $b^{(1)}_{ijk\ldots}$ with $p$ and $\sqrt{T_{NEMS}/T_0}$.
It is then difficult to conclude anything about the geomety-and-temperature independence of $\alpha$. However, it is quite probable that it means that {\it all}  $b^{(1)}_{ijk\ldots}$ are constant; even though higher orders $b^{(n)}_{ijk\ldots}$ $n>1$ are not. This is why we speculate on a mechanism ''driven by the physical mismatch that the surface introduces in the problem, regardless of the {\it excitations} that it can support'', and why we propose that $\alpha$ (or the $b^{(1)}_{ijk\ldots}$) represents a {\it universal feature} of the Knudsen layer.

Even if the correspondence between the expansions of $\delta n$ and $K_{near\,field}$ is not mathematically exact, we can then phenomenologically analyze the damping force as resulting from an {\it effective density} felt by the NEMS as it gradually enters the Knudsen layer, when $\lambda/g$ grows. This is what we do in the core of the paper, for low pressure data, applying Eq. (\ref{molec}) and recalculating from the measured damping the effective gas density $n_{ef\!f}$ replacing $n$ in $\rho_g = m_{H\!e} \, n$. Note that in this range, the expansions in $\lambda/x$ are invalid and no strict correspondence between density and force can be inferred anyway; we thus simply claim that the effective density fit on the data has features that prove the importance of surface excitations: i.e. the anomalous pressure dependence ($\propto P^2$) and fast change with $T$ around the adsorption energy. This shall be in-built in the $b^{(n)}_{ijk\ldots}$ for $n>1$, when $x$ is far enough from 0 for the expansions to be valid.\\

$\,$

$\,$

$\,$

$\,$

$\,$

$\,$


\begin{thebibliography}{blaaaaaaaaaaaaaaaaaaaaaaaaaaaaaaaaaaaaaaaaaaaa}


\bibitem{Carlo} Carlo Cercignani, {\it Mathematical methods in kinetic theory}, Springer Science+Business Media, New York (1969).
\bibitem{Chapman} S. Chapman and T.G. Cowling, {\it The mathematical theory of non-uniform gases}, Cambridge University Press, Third Ed. (1970).
\bibitem{grad} H. Grad, {\it On the theory of rarefied gases}, Comm. on pure and applied Mathematics {\bf 2}, 331 (1949).
\bibitem{patterson} G.N. Patterson, {\it Molecular Flow of Gases}, John Wiley \& Sons inc., New York (1956).
\bibitem{Sone} Y. Sone, {\it Molecular gas dynamics}, Birkhauser Boston (2007). 
\bibitem{microChevrier} J. Laurent, A. Drezet, H. Sellier, J. Chevrier and S. Huant, Phys. Rev. Lett. {\bf 107}, 164501 (2011). 
\bibitem{lockerby} D.A. Lockerby, J.M. Reese, D.R. Emerson and R.W. Barber, Phys. Rev. E {\bf 70}, 017303 (2004).
\bibitem{favero} E. Gil-Santos, C. Baker, D.T. Nguyen, W. Hease, C. Gomez, A. Lemaitre, S. Ducci, G. Leo and I. Favero, Nat. Nanotech. {\bf 10}, 810 (2015).
\bibitem{Bocquet} L. Bocquet, J.L. Barrat, Soft Matter {\bf 3}, 685 (2007).

\bibitem{dadzie} S. Kokou Dadzie, J. Gilbert M\'eolans, Physica A {\bf 358} 328–346 (2005).
\bibitem{sader} Charles R. Lilley and John E. Sader, Phys. Rev. E {\bf 76}, 026315 (2007).
\bibitem{saderII} Charles R Lilley and John E Sader, Proc. R. Soc. A {\bf 464}, 2015 (2008). 
\bibitem{scientreports} T. Wu and D. Zhang, Scientific Reports {\bf 6} 23629 (2016).

\bibitem{Maxwell} J.C. Maxwell, Phil. Trans. Royal Soc. London {\bf 170}, 231 (1879).

\bibitem{NIST} NIST technical note 1334, {\it Thermophysical properties of helium-4 from 0.8 to 1500 K with pressures to 2000 MPa}, Vincent D. Arp and Robert D. McCarty (1989).

\bibitem{JLTPKunal} M. Defoort, K.J. Lulla, C. Blanc, H. Ftouni, O. Bourgeois and E. Collin, J. of Low Temp. Phys. {\bf 171}, 731 (2013). 
\bibitem{SensActCleland} A.N. Cleland and M.L. Roukes, Sensors and Actuators {\bf 72}, 256 (1999).
\bibitem{RSIus} E. Collin, M. Defoort, K. Lulla, T. Moutonet, J.-S. Heron, O. Bourgeois, Yu. M. Bunkov and H. Godfrin, Rev. Sci. Instrum. {\bf 83}, 045005 (2012).
\bibitem{SM} Supplementary Material; contains details on the experimental setup and calibrations. Presents also the mathematical treatment based on both the Grad and Chapman-Enskog methods. 

\bibitem{Yamamoto} Kyoji Yamamoto and Kazuyuki Sera, Phys. of Fluids {\bf 28}, 1286 (1985).
\bibitem{PREpaper} Rustom B. Bhiladvala and Z. Jane Wang, Phys. Rev. E {\bf 69}, 036307 (2004).

\bibitem{saderBeam} J.E. Sader, J. of Appl. Phys. {\bf 84}, 64 (1998).
\bibitem{recentEkinci} V. Kara, V. Yakhot, and K. L. Ekinci, Phys. Rev. Lett. {\bf 118}, 074505 (2017).
\bibitem{bullard} Elizabeth C. Bullard, Jianchang Li, Charles R. Lilley, Paul Mulvaney, Michael L. Roukes and John E. Sader, Phys. Rev. Lett. {\bf 112}, 015501 (2014). 

\bibitem{PRLus} M. Defoort, K. J. Lulla, T. Crozes, O. Maillet, O. Bourgeois, and E. Collin, Phys. Rev. Lett. {\bf 113}, 136101 (2014).
\bibitem{siewert} C. E. Siewert, Phys. of Fluids {\bf 15}, 1696 (2003).

\bibitem{nanoEkinci} V. Kara, Y.-I. Sohn, H. Atikian, V. Yakhot, M. Loncar, and K. L. Ekinci, Nano Lett. {\bf 15}, 8070 (2015).


\bibitem{Dash} J.G. Dash, {\it Films on Solid Surfaces}, Academic Press, London (1975).

\bibitem{Pobell} F. Pobell, {\it Matter and Methods at Low Temperatures}, Springer (1996).



\bibitem{DA} M.M. Dubinin, and V.A. Astakhov, Izv. Akad. Nauk USSR, Ser. Khim., 5 (1971). 

\bibitem{Rudzinski} W. Rudzinski, and D.H. Everett, {\it Adsorption of Gases on Heterogeneous Surfaces}, Academic Press, London (1992).

\bibitem{Cerofolini} G.F. Cerofolini, Surface Science, \textbf{24}, 391 (1971).

\bibitem{Hutson} N.D. Hutson, and R.T. Yang, Adsorption, \textbf{3}, 189 (1997).


\bibitem{Schildberg} P. H. Schildberg, PhD thesis (1988), reprinted in H. Godfrin and R.E. Rapp, Adv. in Physics \textbf{44}, 113 (1995).



\end{thebibliography}
\end{document}